\documentclass[hidelinks]{article}
\usepackage[T1]{fontenc}
\usepackage[a4paper]{geometry}
\usepackage{listings}
\usepackage{multicol}
\usepackage[ruled,vlined]{algorithm2e}
\usepackage{algcompatible}
\usepackage{babel}
\usepackage{amsmath}
\usepackage{amssymb}
\newcommand{\specialcell}[2][c]{%
	\begin{tabular}[#1]{@{}l@{}}#2\end{tabular}}

\usepackage[]{authblk}
\usepackage{mathtools}
\usepackage{setspace} 
\usepackage{hyperref} 

\usepackage[none]{hyphenat}

\usepackage[round]{natbib} 

\usepackage{float}
\begin{document}
	
\title{Learning from zero: how to make consumption-saving decisions in a stochastic environment with an AI algorithm
	\thanks{A previous version of this paper was circulated as the CESifo working paper No. 9255. I am grateful for the advice and constant support of Prof. Roger Farmer and Prof. Herakles Polemarchakis. I appreciate the financial support from Warwick University. I would also like to thank participants at the Society for Computational Economics 27th International Conference, CES 2021 annual conference and the 2021 CESifo Area Conference on Macro, Money, and International Finance. All remaining errors are mine.}}

\author{\large Rui (Aruhan) Shi \thanks{PhD candidate at the department of economics,
		University of Warwick. Email: \href{mailto:a.shi@warwick.ac.uk}{a.shi@warwick.ac.uk}.}\\ 
}

\date{This draft: \today}

\maketitle

\thispagestyle{empty}

\begin{abstract}

This exercise proposes a learning mechanism to model economic agent's decision-making process using an actor-critic structure in the literature of artificial intelligence. It is motivated by the psychology literature of learning through reinforcing good or bad decisions. In a model of an environment, to learn to make decisions, this AI agent needs to interact with its environment and make explorative actions. Each action in a given state brings a reward signal to the agent. These interactive experience is saved in the agent's memory, which is then used to update its subjective belief of the world. The agent's decision-making strategy is formed and adjusted based on this evolving subjective belief. This agent does not only take an action that it knows would bring a high reward, it also explores other possibilities. This is the process of taking explorative actions, and it ensures that the agent notices changes in its environment and adapt its subjective belief and decisions accordingly. Through a model of stochastic optimal growth, I illustrate that the economic agent under this proposed learning structure is adaptive to changes in an underlying stochastic process of the economy. AI agents can differ in their levels of exploration, which leads to different experience in the same environment. This reflects on to their different learning behaviours and welfare obtained. The chosen economic structure possesses the fundamental decision making problems of macroeconomic models, i.e., how to make consumption-saving decisions in a lifetime, and it can be generalised to other decision-making processes and economic models.

\end{abstract}



\newpage

\section{Introduction}


In macroeconomic models, an economic agent's decision is often driven by the desire to maximise its utility. The utility-maximising decision depends on the fundamentals of the economy and the agent's preference. With the rational expectation hypothesis \citep{Muth1961, LUCAS1972103, Sargent1972}, a utility-maximising agent is often perceived as too smart given that it knows the underlying economic structure and is able to form model-consistent beliefs. With adaptive learning algorithms\footnote{See, for example, \cite{Bray1982}, \cite{MarcetSargent1989}, \cite{Sargent1993} and \cite{EvansHon2001}.}, an agent does not know the actual parameter values of the economic structure. Similar to an econometrician, it updates these unknown parameters by running a regression. How an economic agent learns to form a model-consistent belief, or why an econometric learning agent follows a particular regression equation or a decision rule are seldomly discussed. This exercise models bounded rational agent from an angle that is motivated by the psychology literature of learning through reinforcing good or bad decisions. It circumvents the debate on the particular functional form of a learning rule, and draws inspiration from the recent development in the artificial intelligence (AI) literature. More specifically, I model how an economic agent under an actor-critic structure that is proposed in the AI literature learns to make decisions in a stochastic optimal growth environment. It focuses on modelling how this AI agent learns in an unknown environment when it is not aware of its own preference nor the fundamentals of the economy. It must learn through interacting with the environment. Moreover, its behaviours are adaptive to a constantly changing environment owing to a special `exploration' property of the algorithms. 

AI technologies focus on decision-making of an intelligent entity, and they are widely adopted and successfully implemented in situations that normally require human intelligence, such as visual perception, speech recognition, and translation between languages. At the core of AI technologies is the class of algorithms called deep reinforcement learning (DRL), which takes the middle ground of reinforcement learning (RL) and deep learning (i.e., deep artificial neural networks). RL is motivated by how animals and humans learn in the real world, i.e., learning through reinforcing good/bad decisions based on some reward signals.\footnote{in reality, what differ us humans from a reinforcement learning agent is that rewards given by the reality are often not clear and understandable.} Not only are RL and DRL algorithms being widely applied in computer science and AI research, they are also connected to neural scientific research. \cite{DBLP:journals/corr/abs-2007-03750} argue that RL provides a promising theory to explain neural mechanisms of learning and decision-making. One, perhaps most impactful research thus far, has been the empirical evidence that establishes the link between phasic dopamine release and an RL algorithm reward-prediction error signal \citep{NIV2009139}. 

In this paper, a model of the environment is first determined. This is a representation of how the environment behaves. For example, given a state drawn from a state space, and action determined by the agent, the model of the environment shows what the next state is. An agent's decision-making centre or its brain is modelled by two artificial neural networks. One called an actor network that approximates how the agent acts. The other called a critic network that approximates the agent's subjective belief over expected future returns given a particular action in a state of an environment. The agent learns through interacting with the environment, meaning that it tries out different actions and observe its corresponding reward and what the next state turns out to be. The agent makes explorative actions, and this means that the agent adds a level of randomness in its decision. Similar to how humans learn in real life, this is to ensure that the agent tries out different options in its action space to have a better understanding of which option brings itself a satisfactory reward. Moreover, the agent's action is also linked to how the state transitions to the next. Given these interactive experience for many periods, the agent's actor and critic networks evolve through time. Intuitively, the agent's subjective belief about the world that it lives in evolves as it gains more experience, and it learns to make decisions that produces high (subjective) expected future returns. However, the agent may not learn to make optimal decisions, which is defined by a solution under the rational expectation assumption. This is mainly due to the explorative actions that the agent makes. 

The AI agent's learning characteristics, for example, how much an agent explores, is adjusted to observe the implied heterogeneous behaviours from learning in the same environment. The chosen stochastic optimal growth model is a main building block for many macroeconomic models, and can be generalised to other decision-making processes.

This is not the first time that AI and economics are blended together. \cite{Sargent1993} discusses his agenda in combining AI with macroeconomic modelling. He aims at finding a symmetry between econometricians and economic agents. Giving the agent a learning ability based on a decision rule, the agent, at limits, converges to a rational expectation equilibrium. In his case,  agents behave like professional scientists or econometricians and use methods of scientific inference in collecting information and forming their expectation. He argues that this is an important line of literature because it looks at the transition behaviours and dynamics in a learning process. Earlier than Sargent, Herbert Simon defines the concept of bounded rationality and introduces his approach in adopting AI in making decisions.\footnote{He argues that bounded rationality denotes "the whole range of limitations on human knowledge and human computation that prevent economic actors in the real world from behaving the ways that approximate the predictions of economic theories: including the absence of a complete and consistent utility function for ordering all possible choices, inability to generate more than a small fraction of the potentially relevant alternatives, and inability to foresee the consequences of choosing the alternatives"\citep{Simon2016}.} Different from Sargent, Simon focuses more on the decision-making process rather than its outcome. AI that he suggests is on the heuristic search and problem solving by recognition \citep{Simon2016}.

The combination of AI and economics executed here builds on both Sargent's and Simon's views. The AI agent in this exercise learns first by collecting information through an agent-environment interactive process. This past experience is the foundation of this AI agent's subjective belief about the world, which guides the agent's future decisions. More importantly, the agent's subjective belief evolves as the agent gains more experience.  

In the following sections, I first give an introduction of AI technologies, focusing on DRL and related literature. This is followed by the discussion of a model of an environment and how I build the Al economic agent in this environment. I then present several experiments and corresponding results highlighting the insights of modelling bounded rational agents in the proposed actor-critic structure.

\section{AI Technologies and Reinforcement Learning}
This section gives an overview of reinforcement learning and deep reinforcement learning algorithms. These algorithms are mainly designed for the purpose of optimisation, i.e., solving a learning problem. It is worth noting that the purpose of applying an AI algorithm and the execution of it in this paper is different from its original design. This exercise emphasises the implications when learning agents do not stop exploring their environment.

Artificial intelligence or machine intelligence has been on the centre stage of computer science, and subsequently technological advancement for decades. The field of artificial intelligence, or AI, attempts not just to understand but also to build intelligent entities \citep{RussellPeter2020}. It involves a wide range of machine learning techniques\footnote{Machine learning is about learning from data and making predictions and/or decisions. It is broadly categorised as supervised, unsupervised, and reinforcement learning. In supervised learning, there are labelled data; in unsupervised learning, there are no labelled data. In reinforcement learning, in contrast to supervised learning and unsupervised learning, there are evaluative feedback (i.e., reward signals), but no supervised labels.}.

At the core of AI technologies is the class of algorithms under DRL, which is the combination of RL and artificial neural networks (ANNs). ANNs are used as function approximators, which help RL to deal with environment settings with high-dimensional state and action spaces\footnote{An example is learning to play video games directly from raw pixels.}. Notable developments include teaching AI agents (using DRL algorithms) to play Go and Atari games, and to learn speech recognition.

%

\subsection{Reinforcement Learning: a primer}
\label{RLprimer}
The early history of reinforcement learning (RL) has two main threads that were pursued independently before intertwining in modern RL. One thread concerns learning by trail and error, which originates in the psychology of animal learning. The second thread, which is familiar to most computational economists, concerns the problem of optimal control and its solution using value functions and dynamic programming. For the most part, this thread does not involve any learning. The threads come together in late 1980s \citep{SB2018}.


Modern RL\footnote{For a comprehensive review, please see \cite{SB2018}.} contains a series of algorithms aiming at solving Markov Decision Processes. It is distinguished from other computational approaches by its emphasis on learning by an agent from direct interaction with its environment, without requiring exemplary supervision (e.g., supervised machine learning) or complete models of the environment (e.g., dynamic programming). RL uses the formal framework of Markov Decision Processes to define the interaction between a learning agent and its environment in terms of states, actions, and rewards.
\vspace{1.5cm}
\begin{figure}[H]
	
	\caption{The agent-environment interaction in a reinforcement learning setting}
	\centerline{\includegraphics[width=12cm,height=5cm]{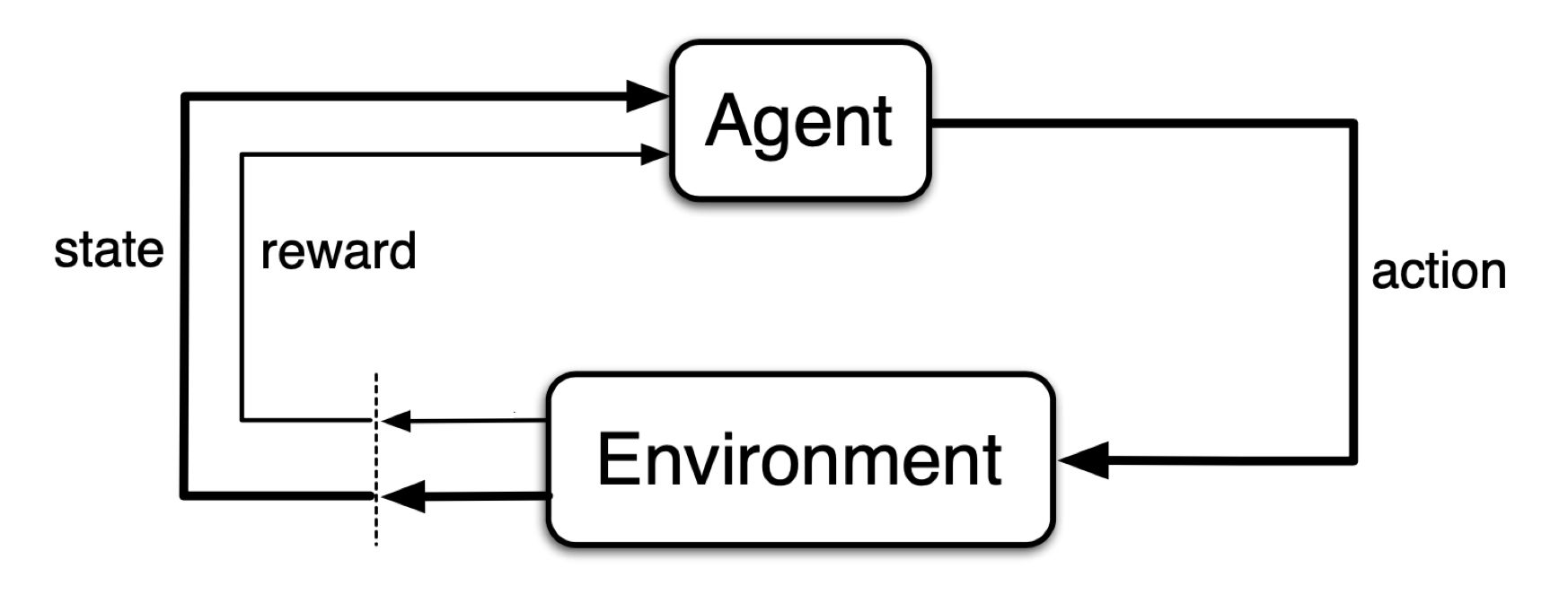}}
	\label{fig1}
	Source: \cite{SB2018}
\end{figure}

\vspace{1.5cm}
An MDP, as described by figure \ref{fig1}, shows a process where given a state variable agent interacts with the environment and chooses an action, this leads to a reward signal for the agent and the current state transits to the next.

At each time step $t$, an RL agent receives some representation of the environment's state (a random variable) out of a state space, $s_t \in \mathcal{S} $, and on that basis selects an action out of an action space, $a_t \in \mathcal{A}(s_t)$. One time step later, in part as a consequence of its action, the agent receives a numerical reward based on a reward function, $r_{t} = r(s_t, a_t)$, and finds itself in a new state, $s_{t+1}$. The new state then feeds into another loop of the agent-environment interactive process. To describe how the state transits, a three-argument function $p :\mathcal{S} \times \mathcal{S} \times \mathcal{A} \rightarrow [0,1]$ is defined as 

\begin{equation}
p(s_{t+1} |s_t, a_t) \equiv Pr\{s_{t+1}|s_t, a_t\}
\end{equation}

for all $s_t, s_{t+1} \in \mathcal{S}$ and $a_t \in \mathcal{A}(s_t)$. It shows the probability of transition to state $s_{t+1}$ from state $s_t$, taking action $a_t$.

A RL agent's task is to learn to make a decision that maximises its expected returns. This expectation is a subjective belief of this agent, and needs not to be based on the true probabilities of the underlying processes. The decision-making strategy, or the RL agent's behaviour is defined by a policy. If an agent follows a stochastic policy $\pi$ at time $t$, then $\pi(a_t|s_t)$ is the probability of choosing action $a_t$ given a state $s_t$. If an agent is following a deterministic policy $\mu$ at time $t$, then $\mu(s_t)$ gives an action for a realised state $s_t$. Expected returns are described by a value function, which estimates how good it is for an agent to perform a given action in a given state \citep{SB2018}. Formally, it is denoted as $Q^\mu(s_t, a_t)$, which shows the expected return after taking an action $a_t$ in state $s_t$ and thereafter following policy $\mu$\footnote{It can also be defined in terms of a stochastic policy $\pi$. Given that the algorithm adopted later follows a deterministic policy, here only value function in terms of a deterministic policy $\mu$ is introduced.}. 

\begin{equation}
Q^\mu(s_t, a_t) \equiv \tilde{E}[R_t|s_t, a_t]
\label{Qpolicy}
\end{equation}

where $R_t = \sum_{k=0}^{\infty} \beta^k r_{t+k}$, and it is the sum of discounted future rewards. $\beta \in [0,1]$ represents a discount factor. $\tilde{E}$ means to show that the expectation is based on an evolving subjective belief of this agent. Equation \ref{Qpolicy} also follows the bellman recursive relationship, 

\begin{equation}
Q^\mu(s_t, a_t) = r(s_t, a_t) + \beta \tilde{E} Q^\mu(s_{t+1}, a_{t+1}),
\end{equation}

where $a_{t+1} = \mu(s_{t+1})$. $\tilde{E}$ denotes a subjective belief.

An Optimal action-value function, defined as 

\begin{equation}
Q^*(s_t, a_t) \equiv \max_{\mu}Q^\mu(s_t, a_t),
\end{equation} 

for all $s_t \in \mathcal{S}$ and $a_t \in \mathcal{A}(s_t)$. For the state - action pair $(s_t,a_t)$, this function gives the expected return for taking action $a_t$ in state $s_t$ and thereafter following an optimal policy.

%
%

%
%

The central assumption in RL is that its agent does not know how a reward $r_t$ is generated, i.e., it does not know the functional form of a reward function. In other words, it does not know how much it likes a certain choice in its choice set. The agent gets to know Its own preference through trying out different options and observe the respective rewards. The agent also does not know the true probabilities, i.e., $p(s_{t+1} |s_t, a_t)$ in its environment. It can only form a subjective belief based on its past experience, and use this belief to guide its future decision. The job of any RL agent is to learn about $r$,  $p$, and $Q$ as best as possible, so as to come up with a decision-making strategy, i.e., a policy $\pi$ or $\mu$, that maximises $Q$ in states of relevance. This process gives a natural machinery to model bounded rationality, as argued by \cite{Abel2019}.

\subsubsection{Exploration vs Exploitation}
To solve a RL problem, one of the challenges that arises is the trade-off between exploration and exploitation. To obtain a lot of rewards, a RL agent must prefer actions that it has tried in the past and found to be effective in producing rewards. But to discover such actions, it has to try actions that it has not selected before. The agent has to exploit what it has already experienced in order to obtain reward, but it also has to explore in order to make better action selections in the future. The dilemma is that neither exploration nor exploitation can be pursued exclusively without failing at the task. The agent must try a variety of actions and progressively favour those that appear to be best. On a stochastic task, each action must be tried many times to gain a reliable estimate of its expected reward.\footnote{Different RL algorithms have different ways of adding exploration.} In reality, we learn about what we like or dislike through trying out different options, which is similar to how learning is modelled in this paper. 

\subsection{Deep Reinforcement Learning}

To ensure RL algorithms can cope with large state space and non-linear value and policy functions, ANNs are combined with the RL algorithms. The resulting class of algorithms are called deep reinforcement learning\footnote{DRL algorithms have already been applied to a wide range of problems, such as robotics, where control policies for robots can now be learned directly from camera inputs in real world, succeeding controllers that used to be hand- engineered or learned from low-dimensional features of the robot's state. In a step towards even more capable agents, DRL has been used to create agents that can meta-learn (`learn to learn'), allowing them to generalise to complex visual environments they have never seen before \citep{Arulkumaranetal2017}.}. Deep refers to the use of ANNs. The pioneer DRL algorithm is called deep Q network algorithm \citep{mnih-atari-2013}, which is capable of human level performance on many Atari video games using unprocessed pixels for input. However, while deep Q network algorithm solves problems with high-dimensional state spaces, it can only handle discrete and low-dimensional action spaces. Many tasks of interest have continuous (real valued) and high dimensional action spaces, including economic decision-making processes. To solve this issue, \cite{lillicrap2015drl} introduces deep deterministic policy gradient (DDPG) algorithm, which is also the algorithm that inspired the learning structure proposed in this exercise.

\subsection{Applications of AI Technologies in Economics}

The literature on applications of DRL in economics is scarce. A majority of the existing literature focuses on other machine learning methods. For example, \cite{NBERc14009} has a survey on the adaptations of machine learning techniques in economics with a particular focus on how machine learning can be used to enhance existing econometric methods. Deep learning (please note that deep learning is a component of but not the same as DRL) is used in stock market predictions. \cite{Minh2018} presents a framework for forecasting stock prices movements concerning financial news and sentiment dictionary. In another study, \cite{Gohong2019} employ deep learning technique to forecast stock value streams while analysing patterns in stock price. Other applications of deep learning include but not limited to fraud detection in insurance industry, auction design, anti-money laundering in banking and online market. Deep learning is also adopted in forecasting macroeconomic indicators but these approaches require huge amounts of data and suffer from model dependency \citep{Mosavi_2020}. \cite{Maliar2019} adopt deep learning to approximate Bellman function and then use supervised learning to train the neural network. \cite{Azinovic2020} apply deep neural networks to solve models with heterogeneity. \cite{Fernández-Villaverde2020} also apply deep neural networks to solve high-dimensional dynamic programming problems.

It is apparent that most machine learning techniques are used in forecasting and predictions. A growing number of papers focus on using deep learning as a solution method for large-scale and heterogeneous economic models. Very few papers focus on applications of DRL algorithms in economics.  \cite{charpentier2020} provide some economic frameworks that could be applied with DRL techniques. This ranges from economic modelling to possible applications in operations research and game theory. They advocate that economic and financial problems would benefit from being reviewed using DRL techniques.  \cite{Shi2021deep} adopt a deep reinforcement learning algorithm to solve a monetary model.

\section{Methodology} \label{Methodology}

In this section, I first show how the environment is usually modelled from the perspective of an economist. I then present and discuss how it is modelled in the setting that an AI learning agent learns to make consumption-saving decisions.

\subsection{The Model: an economist's approach}

In a closed economy with one consumption/capital good, a representative consumer aims at maximising its lifetime utilities:
\begin{equation}
\max_{\{c_t, k_{t+1} \}_{t=0}^{\infty}} E_0\sum_{t = 0}^{\infty} \beta^t u(c_{t})
\end{equation}

subject to 

\begin{equation}
c_t + k_{t+1} = z_t y_t
\end{equation}
\begin{equation}
y_t = k_t^{\alpha}
\end{equation}

\begin{equation}
c_t \geq 0
\end{equation}
\begin{equation}
k_{t+1} \geq 0
\end{equation}
for all $t$.

$c_t$, $k_t$, $y_t$ are consumption, capital investment, and output produced in period $t$ respectively. In this exercise, I use capital investment and saving interchangeably. Period utility $u()$ is increasing and strictly concave, i.e., $u'>0$, and $u''<0$. $\beta$ is the discount factor. Disturbance to the output, $z_t$, is a stochastic random variable, and takes the following form 

\begin{equation}
z_t = e^{\mu + \rho ln(z_{t-1}) + \epsilon_t }
\label{SP}
\end{equation}

where $\epsilon_t$ takes a normal distribution, $\mu$ is a constant, and $\rho$ is an autoregressive parameter.

I take a specific example of the stochastic optimal growth model with logarithmic utility and no capital depreciation. This specification is not a good representation of the real world, nor is it a model for policy experiments. However, it contains the central decision-making problem in economics, i.e., how to make consumption-investment decisions over a lifetime. As the foundation for many popular macroeconomic models\footnote{To name a couple, real business cycle model, and incomplete market model.}, it is a natural starting point to show AI implementation in economic modelling. Moreover, this specification has an analytical solution, which can be used to show how an AI agent generates different and interesting behaviours compared to its rational expectation counterpart formulated by the analytical solution.

\subsubsection{Optimisation under Rational Expectation}
The bellman equation of this problem is as follows.
\begin{equation}
v(k_t, z_t) = \max_{k_{t+1} \in \Gamma(k_t) } \{log(z_t k_t - k_{t+1}) + \beta E_t v(k_{t+1}, z_{t+1})\}
\end{equation}

The solution\footnote{See appendix for detailed derivation.} to this problem is:

\begin{equation}
k_{t+1} = \alpha \beta z_t k_t^{\alpha}
\label{REso}
\end{equation}

The value function following this policy is

\begin{equation}
\begin{split}
v^*(k, z) & = \frac{1}{1-\beta} \left[log(1-\alpha \beta) + \frac{\alpha \beta}{1- \alpha\beta} log\alpha \beta + \frac{\beta \mu }{(1-\alpha\beta)(1-\beta \rho)}\right]  \\
				& + \frac{\alpha}{1 - \alpha \beta} log k + \frac{1}{(1 - \alpha \beta)(1-\beta \rho)} log z \\
\end{split}
\end{equation}

\subsection{The Model: an AI approach}
Inspired by the actor-critic structure of a DRL algorithm\footnote{The algorithm is largely based on the Deep Deterministic Policy Gradient (DDPG) algorithm that is first introduced by \cite{lillicrap2015drl}}. To show how an AI agent learns to make consumption-saving decisions in a stochastic environment, I first show components of RL algorithms, which are introduced in section \ref{RLprimer} and their equivalent representation in a consumption-saving environment, which are presented in table \ref{RL1}.

First, a bounded and compact state space is defined, which represents the world that this AI agent lives. An action space is also defined, which shows the choice set of an agent. In a consumption-saving decision making environment, the state is represented by the total resource available each period, and it is a random variable sampled from the state space. This is presented in the first row of table \ref{RL1}. The agent makes consumption-saving decisions, and the action of the agent is to choose the proportion of the total resource available that it wishes to consume (or save), as shown in the second row of table \ref{RL1}. The action of the agent is also a random variable sampled from the action space. Once an action is made, the agent receives a reward. The reward is determined based on the state and action of a particular period, and it acts as a stimulus signal to show if the agent likes or dislikes a particular choice of action in a given state. The next state is a combination of whatever saved plus the new stochastic income. Naturally, if the agent chooses to consume all in the previous period, it risks zero consumption this period led by no income given the stochastic nature of the income process. In the case of zero consumption, the agent receives minimum rewards (because he would not be happy to be hungry). This dynamic process of interacting with the world helps the agent to form a subjective belief, which guides how the agent acts. What classifies the agent's decision-making centre (or brain) involves two artificial neural networks (the last two rows of table \ref{RL1}). One actor network maps a realised state to an action, whereas the other network called critic approximates the agent's subjective belief of expected returns given a state and an action. Both networks are randomly initialised.



\begin{table}[H]
	\centering	
	\caption{RL components and the economic environment}	
	\label{RL1}	
	\begin{tabular}{@{}lll@{}}\\ \hline \hline				
		\textbf{Terminologies} & \textbf{Description} & \textbf{\specialcell{Representation in the\\ economic environment}} \\ \hline		
		\vspace{0.5cm}		
		\textbf{State, $s_t$}  & \specialcell{A random variable from a state space, \\ $s_t \in \mathcal{S}$} & \specialcell{total goods available\\to consume at period $t$, $z_t k_t^{\alpha}$}   \\
		\vspace{0.5cm}		
		\textbf{Actions, $a_t$}  & \specialcell{A random variable from an action space, \\ $a_t \in \mathcal{A}$} &  \specialcell{proportion of the total goods\\that the agent is willing\\to consume at period $t$}\\	
		\vspace{0.5cm}		
		\textbf{Rewards, $r_t$} & A function of state and action  & \specialcell{utility at period $t$, $ln(c_t)$\\ where $c_t = a_t z_t k_t^{\alpha}$}\\	
		\vspace{0.5cm}		
		\textbf{Next State, $s_{t+1}$}& A random variable from a state space & \specialcell{total goods available\\to consume at period $t+1$,\\where $k_{t+1} = (1 - a_t) z_t k_t^{\alpha}$ }\\ 
		\vspace{0.5cm}
		\textbf{\specialcell{Policy function,\\ $\mu(s|\theta^\mu)$}} & \specialcell{A mapping from state to action, \\ $\mu: \mathcal{S} \rightarrow \mathcal{A}$} & \specialcell{Approximated by a neural network,\\ ie., actor network; \\ parameterised by $\theta^\mu$ \\to be updated during learning}  \\
		\vspace{0.5cm}
		\textbf{\specialcell{Value function, \\$Q(s,a|\theta^Q)$}} &\specialcell{the `expected' (subjective belief) \\return of taking an action in a state}  &\specialcell{Approximated by a neural network,\\ ie., critic network; \\parameterised by $\theta^Q$ \\to be updated during learning}\\ \hline \hline
	\end{tabular}
	
\end{table}

\vspace{1.5cm}		

An AI agent does not know what form of preference it has, nor the fundamentals of the economy. It must gather these information by taking an action, i.e., making a consumption-investment decision, each period. How it decides what action to take given each state depends on its policy function, approximated by a neural network, called actor. More specifically, the algorithm maintains a parameterised actor network $\mu(s|\theta^\mu)$, which specifies the current policy by deterministically mapping states to a specific action given some parameter $\theta^{\mu}$. Given that an AI agent, at the beginning of a learning process, knows nothing about what action constitutes a high reward and lifetime utilities\footnote{Lifetime utilities, i.e., the value function, is approximated by the other neural network, namely critic network.}, it has to try many different actions at each state to have a good idea of what works best. This depends crucially on the agent's ability to explore its action space. 

%

To make sure that the agent is exploring its action space, an exploration policy $\mu'$ is constructed by adding a noise process $\mathcal{N}$
to the actor policy

\begin{equation}
\label{OU}
\mu'(s_t) = \mu(s_t|\theta^\mu) + \sigma_t \mathcal{N}_t.
\end{equation}

$\mathcal{N}_t$ is sampled from a discretised Ornstein-Uhlenbeck (OU) process.\footnote{This noise could be sampled from an uncorrelated Gaussian process or a correlated Ornstein-Uhlenbeck (OU) process. It is likely that a learning agent explores its environment in a correlated manner, and thus a OU process is followed.} $\sigma_t$ represents the scale of impact of this noise $\mathcal{N}_t$, and it represents how explorative an agent is. In theory, it can take any value between 0 and 1. It can also vary through time. For example, at the beginning of a learning process, the exploration is likely to be high since the agent needs to try out more options to gain some level of information. As the agent learns more, it knows what action could bring high reward, and thus may reduce its level of exploration. In this exercise, $\sigma_t > 0$ for all $t$, and this means that the agent always explores. In a non-stationary environment, or an environment that subjects to changes, this implies that the agent keeps its eyes open and can notice the point that an action in a state does no longer bring high reward (due to a change in the environment). Therefore, the agent can adapt to this and learn to make decisions in the new environment.

What constitutes a good policy function depends on the expected return of following a policy. This expected return is determined by the value function, i.e.,$Q^\mu(s,a)$. Assume there are two policies, $\mu_1$ and $\mu_2$, $\mu_1$ is more desirable if $Q^{\mu_1}(s,a|\theta^Q) > Q^{\mu_2}(s,a|\theta^Q) $. That is to say, given the same realised state and action pair, the better policy produces higher expected return, which is measured by the value function $Q$. The value function is also approximated by a neural network, called critic\footnote{Similar to an inner critic we might hear when making a decision.}. As the critic network approximates expected returns, this expectation is contained in the evolving parameter of the critic network $\theta^Q$. This subjective expectation evolves over time as the agent experience more in an environment.


\subsubsection{Full Algorithm and Sequence of Events}
The full algorithm and the sequence of events follow three main steps:
\vspace{1cm}

Step I: Initialisation

\begin{itemize}
	\item Set up the model of an environment. This includes a bounded and compact state space; a bounded and compact action space; a reward function that takes the argument of two random variables (state and action) and outputs a numerical value that represents the reward; state transition dynamics, i.e., an evolution of how current state and action lead to the next state.
	\item Sep up two neural networks: an actor network $\mu(s|\theta^\mu)$ takes the argument of state and output an action; a critic network $Q(s,a|\theta^Q)$ takes the argument of a state-action pair and output a value. 
	\item $\theta^\mu$ and $\theta^Q$ represent the parameters of the two networks respectively. Both are initialised randomly. Both parameters update during the learning process so that the networks will move towards the true policy and value functions.
	\item Define a replay buffer $\mathcal{B}$, which is a memory that stores information (called transitions in the DRL literature) collected by a DRL agent during the agent-environment interactive process. A transition is characterised by a sequence of variables $(s_t, a_t, r_t, s_{t+1})$.
	\item Define a length of $N$, which is the size of a mini-batch. A mini-batch refers to a sample from the memory. 
	\item Define the total number of episodes $E$.
	\item Define a simulation period of $T$ for each episode, where $T>N$. The higher the episodes, the longer the learning periods.\footnote{In the DRL literature, AL agent is usually set to learn a particular task or an Atari game. An episode, thus, represents re-starting the game or task, and it ends with a terminal state (i.e., the end result of a game). In an economic environment, however, a clear terminal state can be difficult to specify. Therefore, the concept of episodes only correlates to how long an agent has been learning.}

\end{itemize}

For each episode, loop over step II and III.\\

Step II: The AI agent starts to interact with the environment.

\begin{itemize}
	\item The agent observes a state, i.e., a realisation of a random variable, $s_t = z_t k_t^{\alpha} $, which is the total goods available to consume at period $t$. It then selects an action (the proportion of the total goods that it is willing to consume) $a_t = \mu(s_t|\theta^{\mu})+\mathcal{N}_t$ according to current policy (actor network) and an exploration noise.
	\item Execute action $a_t$, and observe a reward, which is derived based on the utility function, i.e.,$r_t = ln(c_t)$ and the next state realisation is, $s_{t+1} = z_{t+1} k_{t+1}^{\alpha}$.
	\item Store a transition $(s_t, a_t, r_t, s_{t+1})$ in the memory $\mathcal{B}$.
\end{itemize}

Step III: Training the AI agent (when the AI agent starts to learn) for period $N \leq t \leq T $.

\begin{itemize}
	\item Sample a random mini-batch of N transitions $(s_i, a_i, r_i, s_{i+1})$ from the memory $\mathcal{B}$.
	\item Calculate a value $y_{i}$ for each transition $i$ following
	\begin{equation}
		y_{i} = r_i + \beta Q^{\mu}(s_{i+1},\mu(s_{i+1}|\theta^{\mu})|\theta^{Q})
	\end{equation}
	for all $i\in N$, where $Q^{\mu}(s_{i+1},\mu(s_{i+1}|\theta^{\mu})|\theta^{Q})$ is a prediction made by the critic network with state-action pair $(s_{i+1},\mu(s_{i+1}|\theta^{\mu}))$, and $\mu(s_{i+1}|\theta^{\mu})$ is a prediction made by the actor network with input $s_{i+1}$. 
	
	%
	\item Obtain $Q(s_i, a_i|\theta^Q)$ from the critic network with input state-action pair $(s_i, a_i)$
	
	
	\item Calculate the average loss for this sample of $N$ transitions
	\begin{equation}
		L = \frac{1}{N}\sum_i\big(y_i-Q(s_i,a_i|\theta^Q)\big)^2.
	\end{equation}
	\item Update the critic network with the objective of minimising the loss function $L$.\footnote{This involves applying back propagation and gradient descent procedures.} 
	
	\item For the policy function, i.e., the actor network, the objective is to maximise its corresponding value function. This means that a value function $Q(s,a|\theta^Q)$ that follows a particular policy. In other words, the input action $a$ of $Q$ function is from the policy $\mu$, $a = \mu(s|\theta^\mu)$. Define the objective function as,
	
	\begin{equation}
		J(\theta^{\mu}) = Q^\mu(s_i, \mu(s_i|\theta^{\mu})|\theta^Q).
	\end{equation}
	
	\item This objective function could also be rephrased as minimising $-J(\theta^\mu)$. Update the actor network parameters with the objective of minimising $-J(\theta^\mu)$.\footnote{Similar to the critic network, the specific steps of updating ANN's parameters by minimising an objective function involves back propagation and gradient descent.}
	
\end{itemize}

\vspace{1.5cm}		

\subsection{Parameters and Learning Agent's Characteristics}

The main parameters are presented in table \ref{param}. 

\begin{table}[H]
	\centering	
	\caption{Main Parameters}	
	\label{param}	
	\begin{tabular}{@{}ll@{}}\\ \hline \hline				
		\textbf{Parameters} & \textbf{Baseline Agent} \\ \hline		
				\vspace{0.2cm}		
		\textbf{Output elasticity of capital $\alpha$} &0.4 \\	
				\vspace{0.2cm}		
		\textbf{Shock location parameter $\mu$} &3.0 \\	
				\vspace{0.2cm}		
		\textbf{Autoregressive factor $\rho$} &0 \\	
				\vspace{0.2cm}		
		\textbf{Discount Factor $\beta$} &0.99 \\
				\vspace{0.2cm}		
		\textbf{Learning Rate $\eta$} &actor network $1e-4$; critic network: $1e-3$  \\
		\vspace{0.2cm}		
		\textbf{Exploration Level $\sigma_t$} & 0.3 \\	\hline \hline
	\end{tabular}
	
\end{table}

Learning rate parameter is used for an ANN in the process of updating weights\footnote{For more information on the purpose of a learning rate parameter, please see Appendix.}. 

Exploration level is measured by the scale of noise $\sigma_t$, as specified in equation (\ref{OU}). In this case, the exploration level reduces over time from the full scale of value $1$, and $0.3$, as shown in the last row of table \ref{param}, is the minimum level of exploration a learning agent has.

The exploration parameter is crucial in this paper, not only because it aids the learning process of the neural networks but also because it adds sophistication in the AI agent's learning behaviour, which opens up an unstudied path in modelling economic agents' expectation formation. It represents how an AI agent gathers information. Different exploration levels also mean that agents can have different past experience living in the same environment. An AI agent can be more or less adventurous in exploring available actions (and what these actions lead to). With a higher exploration level, the agent is `willing' to take actions that it has not previously tested, and thus increase the probability of finding a better action (measured by rewards). However, this could also be risky to the agent and it may be left in a worse place than before. With a low exploration level, an agent is unlikely to try anything new, and may never uncover the state-action pairs that contribute to high rewards. This characteristic also allows the AI agent to be alert of any changes in its environment. If a change occurs, for example the autoregressive parameter in $z_t$ (recall that $z_t = e^{\mu + \rho ln(z_{t-1}) + \epsilon_t }$) changes, an AI agent with the ability to explore will notice such changes and adjust its future actions (and hence policy function) accordingly. This way of information collection and processing echoes with empirical evidence on learning from experience and use-dependent brain by \cite{MalmendierNagel2016}, \cite{D'Acuntoetal2021} and \cite{NBERw29336}.

\section{Experiments and Results} \label{Results}


Through several experiments, this section highlights three key results: 1. Learning from zero, the AI agents can reach a stage\footnote{It may not be the stage of full rationality. For comparisons between AI learning agents and a rational expectation agent, please see section \ref{REvsAI}.}, in which their behaviours facing shocks support the permanent income hypothesis argued by \cite{Friedman1957}. 2. AI agents are adaptive to changes in the environment in real time, and the results in this section provide plausible transition dynamics. 3. When AI agents are different in terms of how much they explore the environment (i.e., how they collect information), their transition behaviours are different facing environmental changes, which leads to welfare distinctions. 

In this section, the changes in an environment are introduced through the stochastic process $z_t$ in the economy, recall equation (\ref{SP})

\begin{equation*}
z_t = e^{\mu + \rho ln(z_{t-1}) + \epsilon_t }.
\end{equation*}

More specifically, I position three AI agents in the same environment with changes in the stochastic process. The agents are different only in terms of how much they explore their environment, i.e., their exploration levels are different. I run the following simulations.
\begin{itemize}
	\item Transitory shock: in an environment with $z_t = e^{0.1+\epsilon_t}$, impose a one-time change to the mean and resulting $z_t = e^{3+\epsilon_t}$. Observe AI agents' consumption behaviours in relation to this transitory change.
	\item Permanent change: shift the stochastic process $z_t$ from $z_t = e^{0.1+\epsilon_t}$ to $z_t = e^{0.1+ 0.7ln z_{t-1}\epsilon_t}$. Observe AI agents' consumption behaviours in relation to this permanent change.
\end{itemize}

\subsection{Learning from Zero}

To illustrate that an AI agent learns from square one, I take the example of a baseline agent, and show the difference of its simulated behaviours before and after learning.

\begin{figure}[H]
	\caption{Simulated consumption paths during learning}
	\centerline{\includegraphics[width=19cm,height=6.5cm]{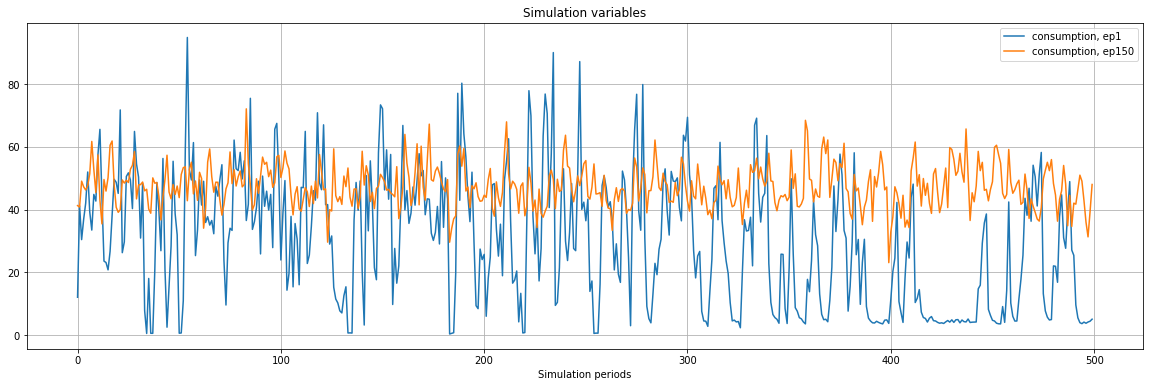}}
	\label{learning_c}
\end{figure}

Figure \ref{learning_c} plots this AI agent's consumption paths at the beginning of a learning process (labelled ep1) and towards the end of a learning process (labelled ep150). The x-axis plots simulation periods. It illustrates that at the beginning of a learning process, this agent's consumption choices (denoted by the blue line) are more volatile than when it has been learning in the environment after many periods (orange line). This shows that the agent does not know what is desirable in its choice set, and thus taking many random actions, which also corresponds to a high exploration level at the beginning of a learning process. After it has been learning in this environment, its decisions are more focused.

\begin{figure}[H]
	\caption{Loss of the policy neural network}
	\centerline{\includegraphics[width=19cm,height=6.5cm]{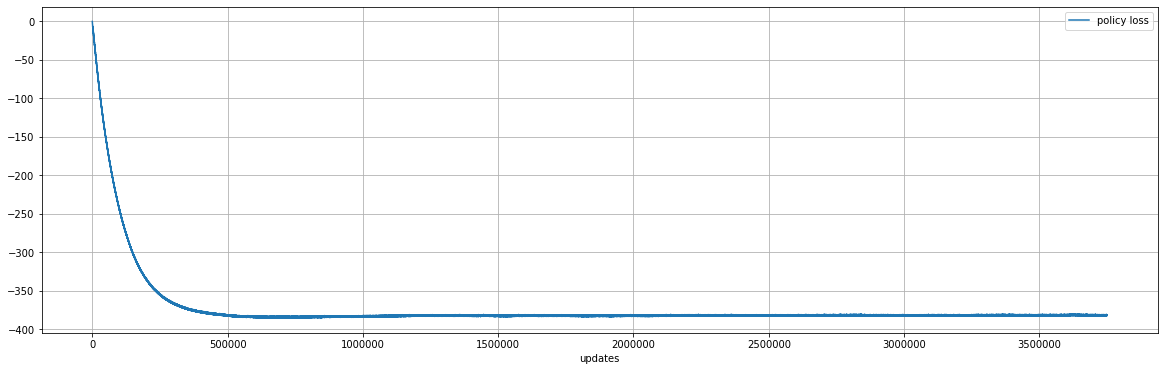}}
	\label{policy_loss}
\end{figure}

Figure \ref{policy_loss} plots the loss of the policy network during this learning process. It plots the gradual reduction of the loss through learning. In other words, the agent makes more decisions that generate high rewards through learning.

In the following subsections, I present results for the simulation experiments with environmental changes. All results are presented for three types of AI agents differing in their exploration levels (i.e., how they collect information), accentuating the importance of exploration parameter in generating different learning dynamics and determining the welfare of AI agents. 

\subsection{Transitory Shock}
\label{trans}
	
	\begin{figure}[H]
	\caption{The stochastic process with a transitory shock}
	\centerline{\includegraphics[width=19cm,height=6.5cm]{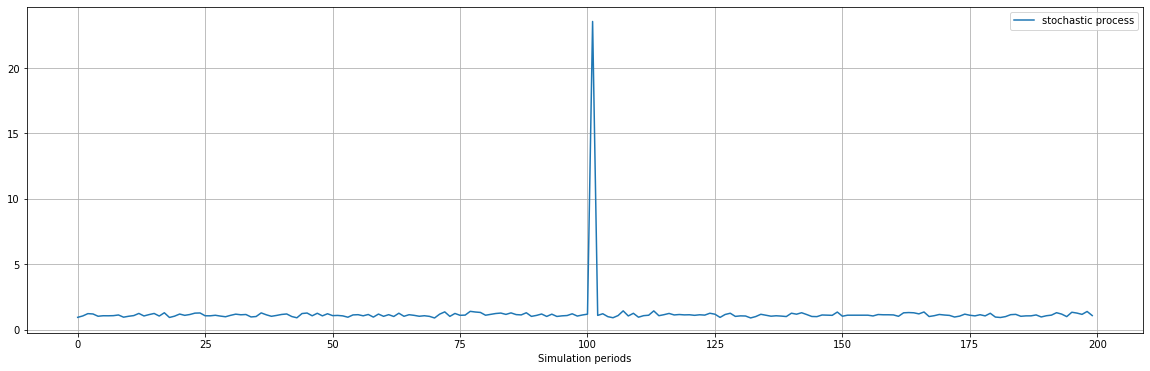}}
	\label{trans endow}
	\end{figure}

\begin{figure}[H]
	\caption{AI agents' consumption paths facing a transitory shock}
	\centerline{\includegraphics[width=19cm,height=6.5cm]{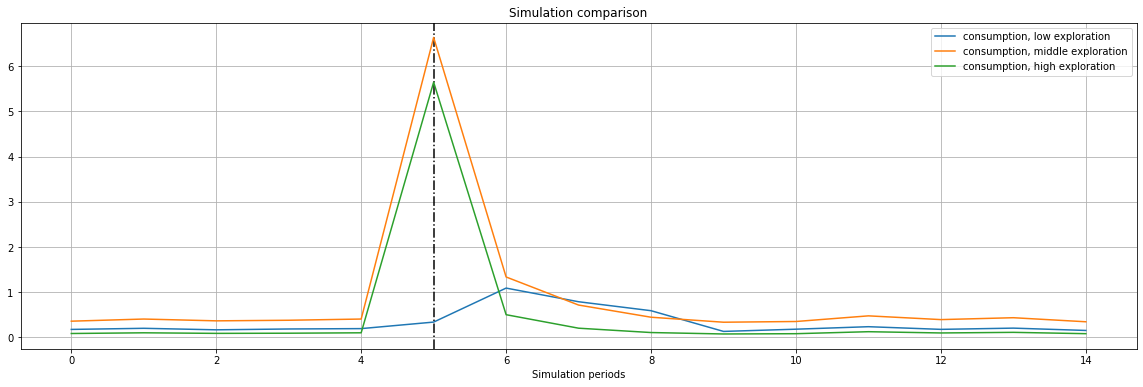}}
	\label{trans c}
\end{figure}

\begin{figure}[H]
	\caption{AI agents' utilities facing a transitory shock}
	\centerline{\includegraphics[width=19cm,height=6.5cm]{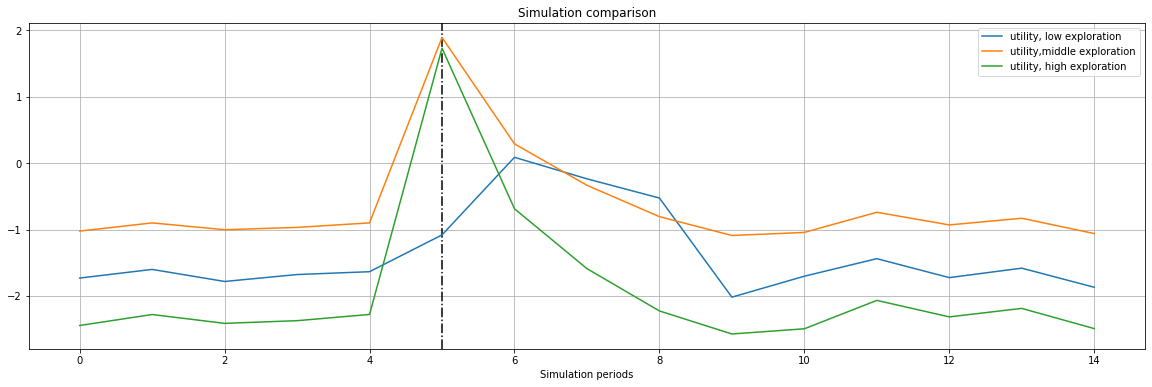}}
	\label{trans u}
\end{figure}

Figure \ref{trans endow} shows the stochastic process in this environment. The process follows $z_t = e^{0.1 + \epsilon_t}$ except for simulation period 100 where $z_t = e^{3 + \epsilon_t}$. $\epsilon_t$ follows a normal distribution with $N(0, 0.1)$. The x-axis represents simulation periods. To clearly show agents' responses, I plot simulation data for the 15 periods around the transitory shock. Hence the x-axis values of figure \ref{trans c} and \ref{trans u} are different from figure \ref{trans endow}. Figure \ref{trans c} plots AI agents' consumption paths in this environment. The transitory shock is unknown to them before it hits. The black vertical dash line represents the period when the positive transitory shock is realised. Before the shock hits, all three agents reach a stage of smooth consumption path through learning. When the shock hits, all three agents with different exploration level exhibit similar overall consumption behaviours, namely consumption increases with the positive shock and revert back to the pre-shock level after a few periods. The timing and magnitude of their responses are different. The middle- and high-exploration agents respond more swiftly than the low exploration agent, which attests that with higher exploration, an agent is more alert to changes in the environment and hence responds quickly. The low-exploration agent responds with a lag, as shown by the blue line. The magnitude of their responses is also correlated to their respective exploration levels. Low-exploration agent responds in a slower and less prominent manner than the other two agents.


%

Figure \ref{trans u} plots all three agents' utility as a measurement of their welfare. The green-line agent (high exploration) does better than the low-exploration agent during the transitory shock period. However, this does not hold for periods before and after the transitory shock. Its overly adventurous nature leads to a behaviour with high excess investment (very low consumption) and thus a lower utility level than the middle agent. The middle agent (orange line) has the highest utility, and balances exploration and exploitation of existing knowledge. If the agent reduces its level of exploration, as shown by the blue line in the figure, it sacrifices its welfare in support of its cautious behaviour and only try actions that it has tested before. 
 
This transitory shock can be interpreted as a positive productivity shock. With a positive productivity shock, a temporary increase in consumption is seen. Interestingly, without any further assumptions, a lagged response can be generated simply through varying AI agents' exploration parameters (i.e., the low-exploration agent in blue in figure \ref{trans c}). 

How would the agents respond in an environment with a permanent change?

\subsection{Permanent Shock}
\label{perm}

\begin{figure}[H]
	\caption{The stochastic process with a permanent change}
	\centerline{\includegraphics[width=19cm,height=6.5cm]{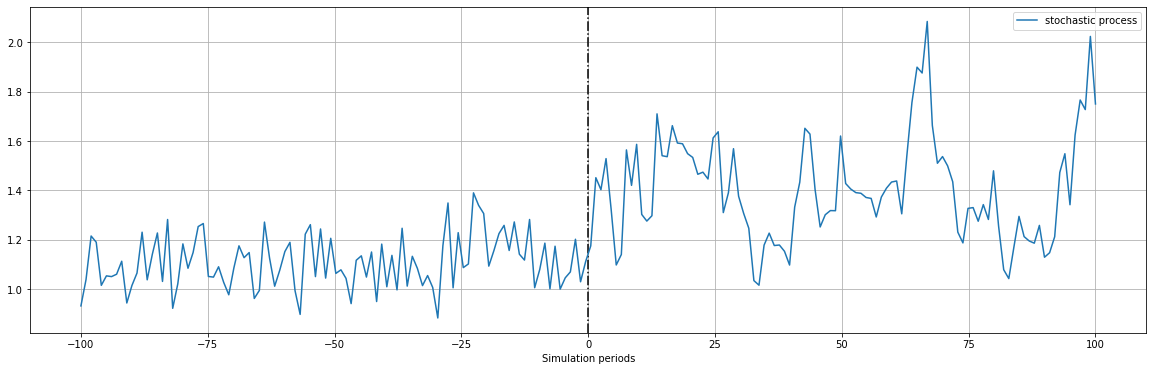}}
	\label{perm z}
\end{figure}

\begin{figure}[H]
	\caption{Learning agent in an environment with a permanent change}
	\centerline{\includegraphics[width=19cm,height=6.5cm]{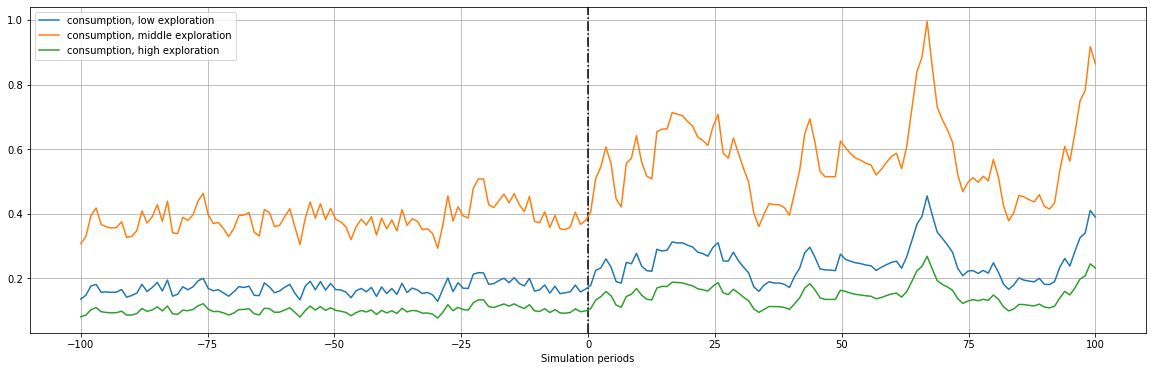}}
	\label{perm c}
\end{figure}


Figure \ref{perm z} shows the stochastic process changes from $z_t = e^{0.1 + \epsilon_t}$ to $z_t = e^{0.1 + 0.7 ln z_{t-1} + \epsilon_t}$. The black dash line indicates the period when the change happens. Given this permanent change of stochastic process, all three AI agents modify their consumption levels permanently, as indicated by figure \ref{perm c}. AI agents' behaviours in environments with transitory and permanent changes echo with Milton Friedman's permanent income hypothesis, and that a permanent income change (rather than a temporary one) drives the change in a consumer's consumption smoothing behaviour \citep{Friedman1957}. 


\begin{figure}[H]
	\caption{Learning agent in an environment with a permanent change}
	\centerline{\includegraphics[width=19cm,height=6.5cm]{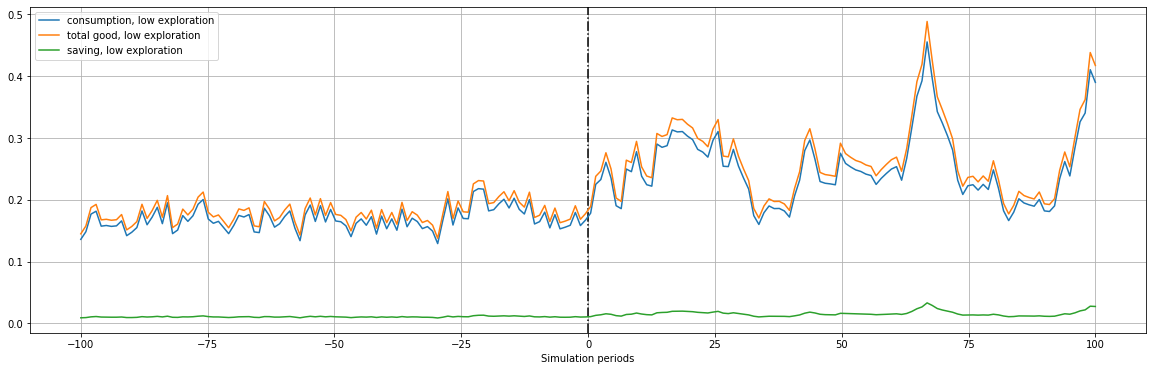}}
	\label{perm L}
\end{figure}

\begin{figure}[H]
	\caption{Learning agent in an environment with a permanent change}
	\centerline{\includegraphics[width=19cm,height=6.5cm]{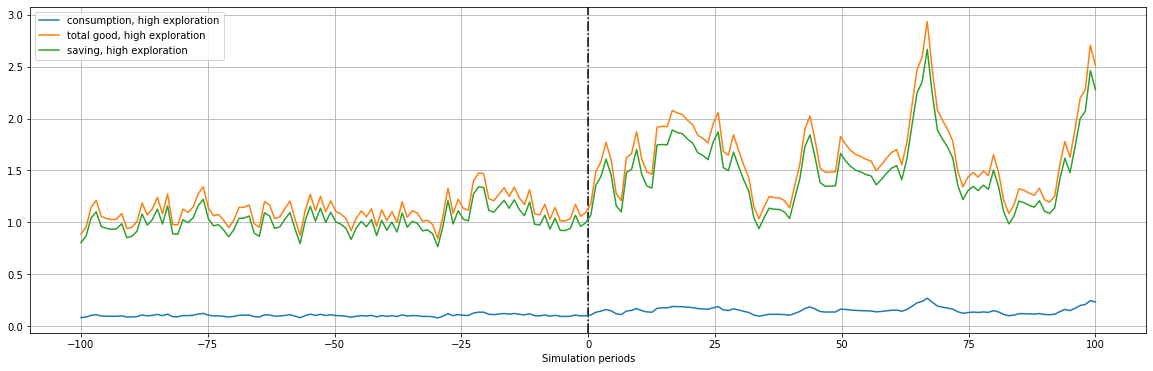}}
	\label{perm H}
\end{figure}

\begin{figure}[H]
	\caption{Learning agent in an environment with a permanent change}
	\centerline{\includegraphics[width=19cm,height=6.5cm]{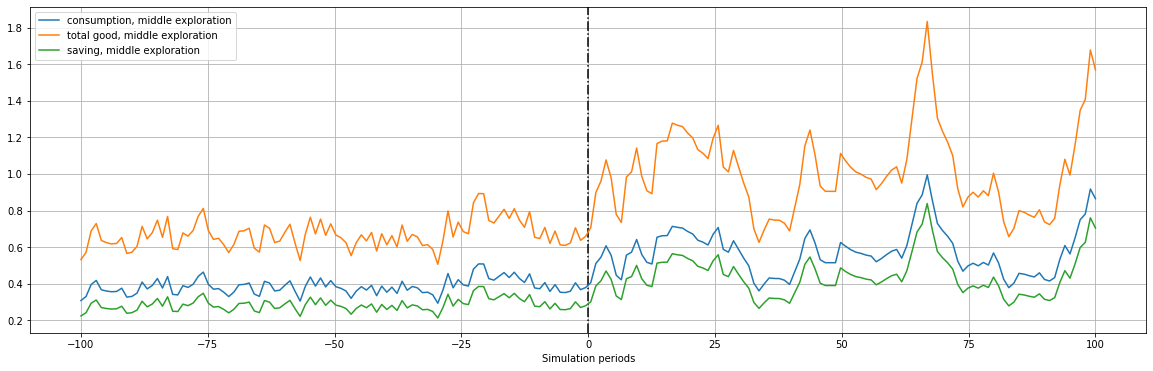}}
	\label{perm M}
\end{figure}

\begin{figure}[H]
	\caption{Learning agent in an environment with a permanent change}
	\centerline{\includegraphics[width=19cm,height=6.5cm]{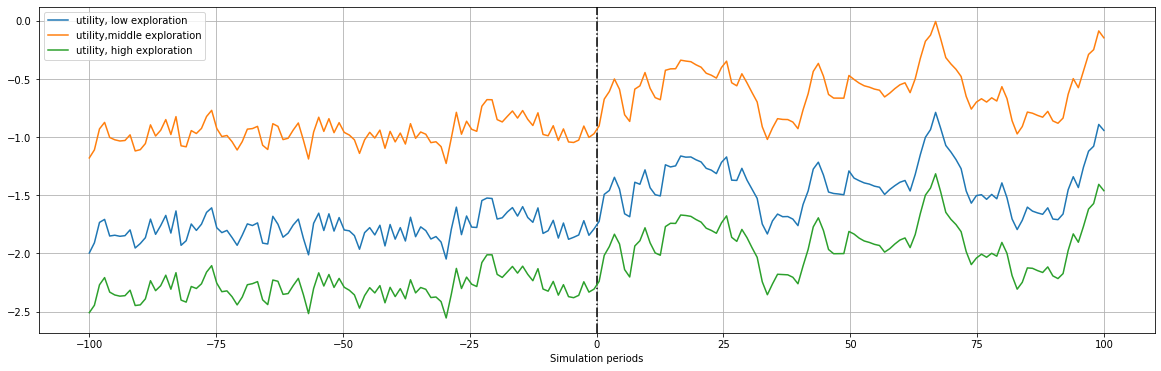}}
	\label{perm u}
\end{figure}

Figure \ref{perm L} plots low-exploration agent's simulated series of consumption, investment, and total available resource each period. As clearly shown, it consumes almost all of its available resource each period. This is reversed in figure \ref{perm H}. The high-exploration agent invests nearly all its available resource each period. The middle-exploration agent takes the middle ground, as shown in figure \ref{perm M}. Their respective behaviours contribute to their welfare distinctions as shown in figure \ref{perm u}.

Figure \ref{perm u} shows all three agents' utility in this environment with a permanent change in the stochastic process. As anticipated, the middle-exploration agent balances exploring the unknown action space and exploiting the known domain to achieve the highest utility. However, the agents with low and high level of exploration sacrifice their welfare to support their overly cautious or adventurous behaviours.

One important issue is that exploration level is a relative term and subjective to a particular environment or a decision-making problem. A high-level exploration in this setting could mean a low-level one in another problem. Therefore, it is important, when applying this technique, to experiment with many different levels of exploration.

%
%

\section{Comparisons with One Agent under Rational Expectation}
\label{REvsAI}

AI agents in this paper are born in an environment that they have no information on what is desirable or feasible in its choice set. It also does not know how state transitions. They first interact with the environment through making decisions and observing the corresponding consequence. This experience is then used to update the agent's subjective belief about the world, which in turn, guides the agent's future decisions. Moreover, to make sure that they get to know available choices given a particular state, the agent must explore, i.e., trying out unknown actions. All these behaviours differ from an economic agent under rational expectation assumption or an econometric learning agent. In this section, I compare AI agent's behaviour with an agent under rational expectation assumption (RE agent). More specifically, I first compare the learnt/approximated policy function with the analytical solution of the given economic model. I then compare the simulated path of the AI agents and the RE agent, and show how their consumption decisions differ.

\begin{figure}[H]
	\caption{Approximated policy functions}
	\centerline{\includegraphics[width=19cm,height=10cm]{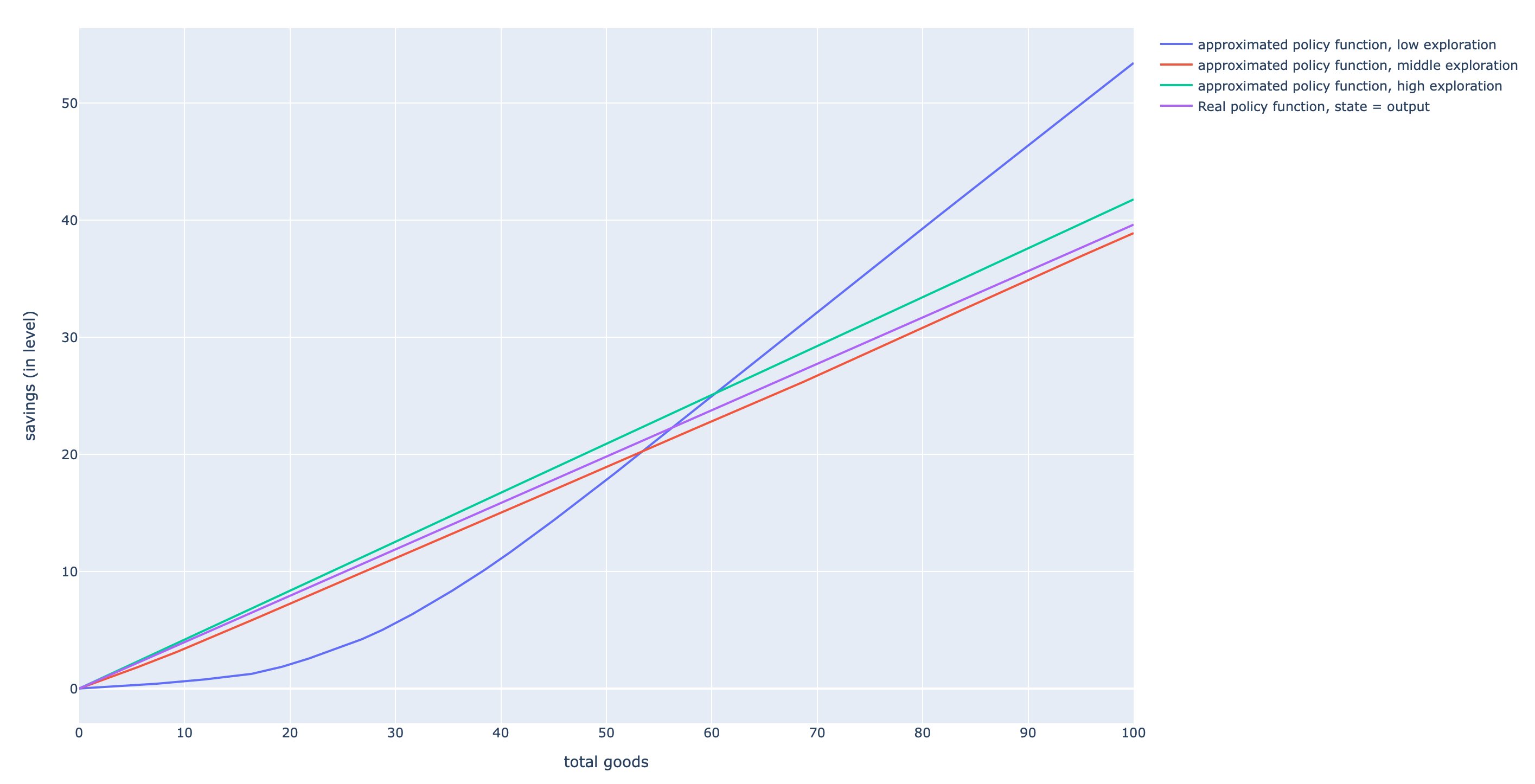}}
	\label{policycomparison}
\end{figure}

\begin{figure}[H]
	\caption{Distance metrics}
	\centerline{\includegraphics[width=19cm,height=10cm]{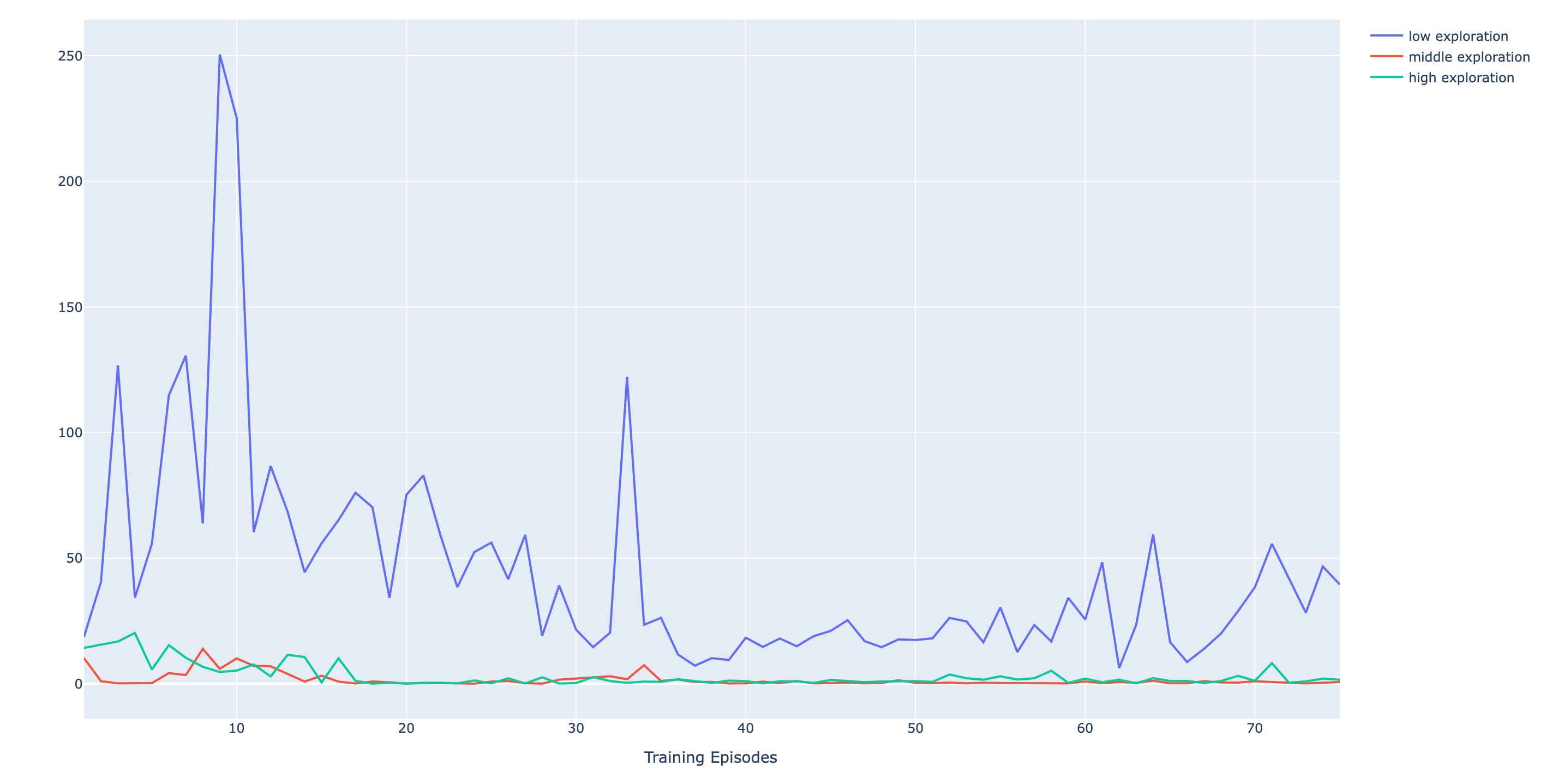}}
	\label{dist}
\end{figure}

Figure \ref{policycomparison} plots the approximated policy functions for three AI agents (with different exploration levels) and the analytical solution policy of this stochastic optimal growth problem (i.e., equation \ref{REso}). All three AI agents have been learning for the same amount of periods. It shows that the AI agents' learnt policy functions can be very close to the solution of the problem with the middle exploration parameter (red line agent). The approximated policy can also be different if the agent is overly cautious or adventurous in choosing their actions (blue and green lines). Figure \ref{dist} plots a distance metrics calculated between the analytical solution policy and the approximated policies at each episode as illustrated by equation (\ref{dista}). $k^*_g$ represents the analytical solution policy function value at a grid $g$, and $k_g$ is the approximated policy function at the same grid. $G$ denotes total number of grids. $d_e$ denotes the distance between analytical solution and approximated policy function at episode number $e$.

\begin{equation}
d_e = \frac{1}{G}\sum_{g=1}^{G}(k^{*}_g - k_g)^2
\label{dista}
\end{equation}

In figure \ref{dist}, the x-axis denotes the number of training episodes, i.e., how long the AI agent has been living and learning in an environment, and the y-axis denotes the distance calculated. It shows that as the number of training episodes increases, the distance becomes smaller. In addition, the middle exploration agent learns the fastest, that is, the distance becomes smaller at earlier episodes than the high and low exploration agents.

\begin{figure}[H]
	\caption{AI agent vs RE agent consumption paths}
	\centerline{\includegraphics[width=19cm,height=6.5cm]{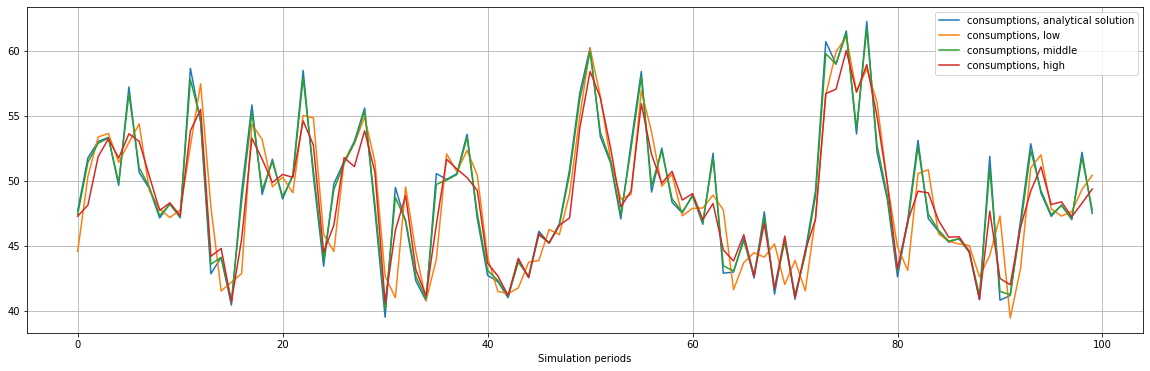}}
	\label{REc}
\end{figure}

%

Figure \ref{REc} plots simulation comparisons among three AI agents and the RE agent in the same stochastic environment. All agents follow the same initial condition and behave following their respective policies in subsequent periods. It shows that the middle exploration agent is doing almost as well as the RE agent in terms of the consumption level, which determines agents' welfare. More importantly, it illustrates that through the learning structure proposed in this exercise, AI agents have the ability to make decisions that are very close to what a rational agent would do, however their decisions need not be identical. This is due to the constant exploration of AI agents.


\section{Summary and Future Work}
In this paper, I show how an economic agent learns to make decisions in an unknown environment and how it adapts to changes in the underlying stochastic process of the economy. Drawing inspirations from the actor-critic structure in the artificial intelligence literature, the economic agent is born without knowing what is feasible and desirable in its choice set. It does not know how state transitions in its environment. Through the proposed learning structure, I model how this agent interacts with its environment and gains experience. The experience is then used to update the agent's evolving subjective belief. The agent's decision-making strategy is formed and adjusted based on its evolving subjective belief. I adopt a version of the stochastic optimal growth model where an economic agent needs to make consumption-saving decisions to maximise its lifetime utilities. This agent, however, does not follow the rational expectation hypothesis. It also does not follow a pre-specified decision or forecast rule similar to econometric learning agents. It is physically constraint to take an optimal action because it does not possess any information on what is feasible or desirable in its choice set. The agent could only learn how to make decisions through acquiring information by interacting with the environment for many periods and processing it through artificial neural networks. 

Several experiments are conducted. After learning in the stochastic optimal growth environment for several periods, to show how AI agents respond to permanent and transitory changes in the environment, I impose a transitory shock and a permanent change to the stochastic process. The AI agents' behaviours echo with the permanent income hypothesis, and that a transitory shock leads to a temporary response from the agent and nothing permanent. Whereas the permanent change in the stochastic process leads to a sustained shifts in agents' consumption behaviours. In addition, to highlight the purpose and novelty of the exploration parameter, I run all experiments on three AI agents differing in their levels of exploration. This leads to differences in their past experience and information collected. With an appropriate exploration level, the agent achieves a high level of welfare measured by its utility. With a higher level of exploration, the agent is overly adventurous, and sacrifices its welfare for an unknown/untested action. With a lower level of exploration, the agent is too cautious and unwilling to try anything unheard of and thus is unable to fully explore and find actions that lead to high rewards. In the end, I show a comparison of policy functions and simulated behaviours between AI agents who have the ability to explore and the rational expectation agent. Their policy functions could be very similar given an appropriate exploration parameter. This is affirmed by their behaviours in a simulated environment.


This work provides a way to model how an economic agent learns to make decisions in an unknown environment, which is motivated by the psychology literature on learning through reinforcing good or bad decisions and the neural science literature of animal learning. It relaxes the rational expectation assumption, and models the learning behaviour of an artificial agent in terms of how it collects and processes information. Differing from an econometric learning agent, AI agents here do not follow a pre-specified learning rule. An AI agent collects information through exploring the environment and processes information through artificial neural networks. This is highly relevant, and supports further studies on important economic questions that include structural breaks, regime changes, and multi-agent learning. Owing to artificial neural networks, the agent's decision-making centre does not need to follow a pre-specified functional form, and it is adaptable to the evolving subjective belief of an AI agent, which echos with the empirical evidence of learning with a use-dependant brain by \cite{NBERw29336}.

\newpage

\bibliographystyle{agsm}
\bibliography{bibtex}

\newpage
\appendix
\section{Appendix}
%


\subsection{Derivation of Analytical Solution of the Stochastic Optimal Growth Model}

Given the bellman equation:
\begin{equation}
v(k_t, z_t) = \max_{k_{t+1} \in \Gamma(k_t) } \{log(z_t k_t - k_{t+1}) + \beta E_t v(k_{t+1}, z_{t+1})\}
\end{equation}

Guess the value function to be the form $v(s) = A + B log(k) + D log(z)$, and substitute to the right hand side of the bellman equation 
\begin{equation}
v(k_t, z_t) = \max_{k_{t+1} \in \Gamma(k_t) } \{log(z_t k_t - k_{t+1}) + \beta E_t (A +B log(k_{t+1}) + D log(z_{t+1}) )\}
\end{equation}

\begin{equation}
v(k_t, z_t) = \max_{k_{t+1} \in \Gamma(k_t) } \{log(z_t k_t - k_{t+1}) + \beta (A +B log(k_{t+1}) + D E_t log(z_{t+1}) )\}
\end{equation}

\begin{equation}
v(k_t, z_t) = \max_{k_{t+1} \in \Gamma(k_t) } \{log(z_t k_t - k_{t+1}) + \beta (A +B log(k_{t+1}) + D \mu )\}
\end{equation}

The F.O.C:

\begin{equation}
\frac{\partial v(k_t, z_t)}{\partial k_{t+1}} = 0 \rightarrow - \frac{1}{z_t k_t - k_{t+1}} + \beta \frac{B}{k_{t+1}} = 0
\end{equation}

\begin{equation}
k_{t+1} = \frac{\beta B}{1+\beta B} z_t k_t
\end{equation}

Apply the envelop theorem 

\begin{equation}
\frac{B}{k_t} = \frac{z_t}{z_t k_t - k_{t+1}} \rightarrow B =  \frac{z_t k_t}{z_t k_t - k_{t+1}}
\end{equation} 

$k_{t+1}$ can then be derived as

\begin{equation}
k_{t+1} = \beta z_t k_t
\end{equation}

If the guessed form of value function is the solution then it must satisfy 

\begin{equation}
A + B log(k_t) + D log(z_t) = log(z_t k_t - \beta z_t k_t) + \beta (A + Blog(\beta z_t k_t) + D\mu)
\end{equation}

\begin{equation}
A = \frac{1}{1-\beta} (log(1-\beta) + \frac{\beta}{1- \beta} log\beta + \frac{\beta \mu }{1-\beta})
\end{equation}

\begin{equation}
B = D = \frac{1}{1-\beta}
\end{equation}

The value function is thus

\begin{equation}
v^*(k, z) = \frac{1}{1-\beta} \{log(1-\alpha \beta) + \frac{\alpha \beta}{1- \alpha\beta} log\alpha \beta + \frac{\beta \mu }{1-\alpha\beta}\} + \frac{\alpha}{1 - \alpha \beta} log k + \frac{1}{1 - \alpha \beta} log z
\end{equation}

When $z_t$ follows an autoregressive process, i.e. $z_t = e^{\mu + \rho ln z_{t-1} + \epsilon_t}$, where $\epsilon_t$ follows a standard normal distribution, the analytical solution is derived analogously.

\begin{equation}
v^*(k, z) = \frac{1}{1-\beta} \{log(1-\alpha \beta) + \frac{\alpha \beta}{1- \alpha\beta} log\alpha \beta + \frac{\beta \mu }{(1-\alpha\beta)(1-\beta \rho)}\} + \frac{\alpha}{1 - \alpha \beta} log k + \frac{1}{(1 - \alpha \beta)(1-\beta \rho)} log z
\end{equation}

Policy function is

\begin{equation}
k_{t+1} = \alpha \beta z_t k_t^{\alpha}
\end{equation}

\subsection{How do ANNs learn?}

\begin{figure}[H]
	\caption{A feedfoward network with one hidden layer}
	\centerline{\includegraphics[width=12cm,height=8cm]{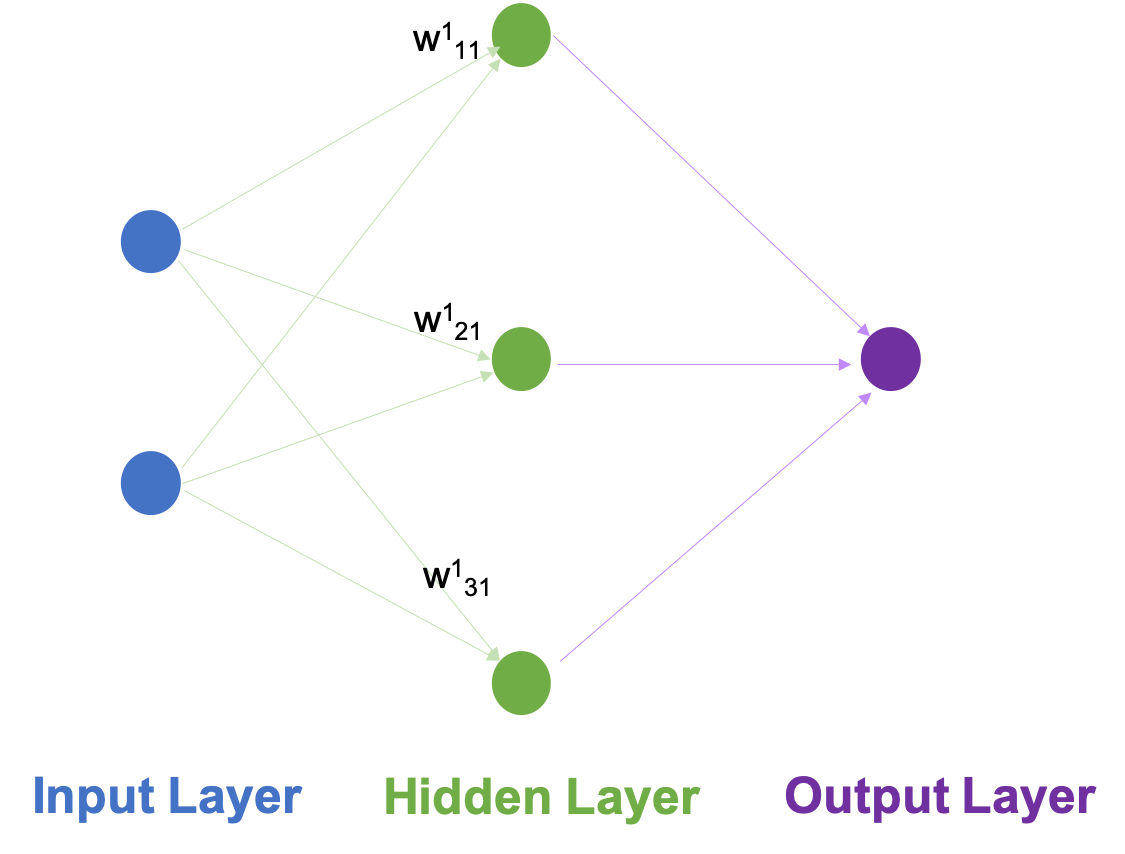}}
	\label{ANN}
	\hspace*{10pt}\hbox{\scriptsize Source: author's own construction}	
\end{figure}

Figure \ref{ANN} shows a feedforward network with one hidden layer. All the circles are neurons (or nodes) of this ANN. The first column is the input layer, and it has two nodes represented by blue. The middle column is the hidden layer, and it has three nodes in green. The last column is the output layer with one node in purple. The arrows represent directions of information/data flow. Feedforward means that the data flows forward from input to output layers. $w$s are weights, which needs to be learnt while training an ANN. The superscripts on $w$s represent the layer that the weights are assigned to. For example, $w^1_{21}$ represents the weight of the first neuron in the input layer (the first layer) to the second neuron in the hidden layer (second layer). 

I use the first node in the hidden layer as an example to show the information flow within a node. Assume that the input layer nodes output $x_1$ and $x_2$ respectively. The first node in the hidden layer then takes information from the previous layer as $w^1_{11} x_1 + w^1_{21} x_2$, and apply an overall bias. The output of this node becomes $\sigma (w^1_{11} x_1 + w^1_{21} x_2 + bias)$, where $\sigma()$ represents an activation function, and it can take many forms (e.g., sigmoid, logistics, and tanh functions). 

How do ANNs learn and update their parameters? It takes the following procedures. Given some training dataset and an ANN model, pass the data forward through the neural network to obtain an output/prediction. Compare this output to a target value/goal. Calculate the loss between the predicted value and the target value. To make sure the neural network learns and update its parameters (including weights of each neuron and biases), a way to link each weight and the loss is needed. \textit{Back propagation} is an algorithm to find such links. In particular, It finds partial derivatives of the loss function with respect to each weight and bias by applying chain rules.

These partial derivatives are called gradients in the literature. Given these gradients, parameters can be updated through gradient methods, such as \textit{gradient descent}. In gradient descent, the update process looks as follows in order to update each weight of the network.

\begin{equation}
w'_{ij} = w_{ij} -\eta\frac{\partial Error}{\partial w_{ij}},
\label{LR}
\end{equation}

where $w_{ij}$ represents weight of the $j^{th}$ neuron to the $i^{th}$ neuron in the next layer. $\frac{\partial Error}{\partial w_{ij}}$ is the partial derivative of the loss function with respect to the weight, and it states how much error the weight $w_{ij}$ contributed. The new weight $w'_{ij}$ is a combination of the current weight $w_{ij}$ and some weighted term $\eta\frac{dError}{dw_{ij}}$. $\eta$ denotes learning rate, a hyperparameter. It represents how quickly weights are updated in response to the error contribution term. The higher the learning rate, the quicker the update is.

\vspace{1cm}
\begin{figure}[H]
	\caption{Learning rates}
	\centerline{\includegraphics[width=12.5cm,height=4.5cm]{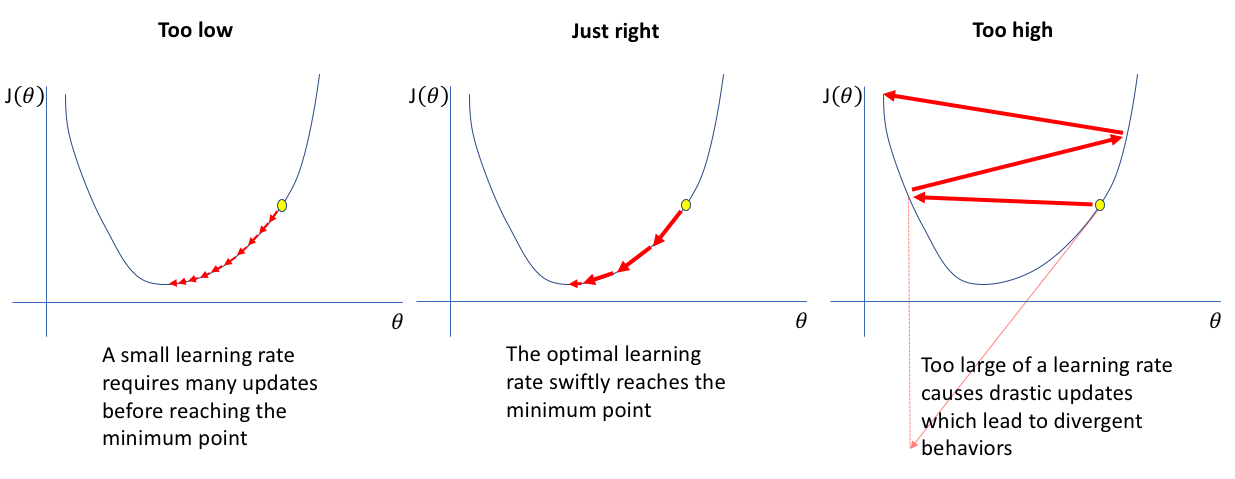}}
	\hspace*{10pt}\hbox{\scriptsize Source: \href{https://www.jeremyjordan.me/nn-learning-rate/}{jeremyjordan}}	
	\label{learningrate}
\end{figure} 
\vspace{1cm}

To clearly illustrate the purpose of $\eta$, figure \ref{learningrate} shows that when the learning rate is too high, it causes a divergence and fails to reach the appropriate weights; with a very low learning rate, the learning process takes very long before it reaches the optimal solution.

The objective of gradient descent is to minimise a loss function. At the point where the loss is minimised, the gradient is 0. Thus, gradient descent requires weights adjustment so as to move along on the line and go to the minimum point, which is similar to what figure \ref{learningrate} illustrates.

%
%
%
%
%
%

\end{document}